\documentclass{appolb}
\usepackage{epsfig}

\begin{document}
\title{A simple model of local prices and associated risk evaluation
}
\author{Krzysztof Urbanowicz
\address{Quant Technology Sp z o.o.\\
Warsaw, 03-186, Poland}
\and
Peter Richmond
\address{School of Physics, Trinity College\\
Dublin 2, Ireland}
\and
Janusz A. Ho{\l}yst
\address{Faculty of Physics, Warsaw University of Technology\\
Warsaw, 00-662, Poland}
}
\maketitle
\begin{abstract}
\par
A simple spin system is constructed to simulate dynamics of asset
prices and studied numerically. The outcome for the distribution of
prices is shown to depend both on the dimension of the system and the
introduction of price into the link measure. For dimensions below 2, the
associated risk is high and the price distribution is bimodal. For higher
dimensions, the price distribution is Gaussian and the associated risk is much
lower. It is suggested that the results are relevant to rare assets or
situations where few players are involved in the deal making process.
\end{abstract}
\PACS{PACS numbers come here}

\section{Introduction}
\par
Frequently one encounters situations where the price of identical
objects varies across sellers. In the US and the UK for example one
can find a number of retailers that proudly display signs stating
'If you can find a lower price elsewhere, tell us and we will match it'.
Searching the internet using engines such as kelco.com allows consumers to
find many different retailers offering identical items at very different prices.
Using a very simple spin model, widely used by other authors, \cite{ref1,ref2,ref3} we give here
some insights into the origin of such situations.

\section{The model and dynamics}
\par
Our model consists of nodes. Each node is a simple spin that
can be either +1 representing a 'buyer' or -1 representing a 'seller'.
The N spins are labeled in sequence from 1 to N and at time t=0 are assigned
the state ±1 randomly. Simultaneously they are each assigned a price, $p_i$ .
A t=0 this is set at zero for all nodes.
\par
The dynamics arises from interactions between pairs of spins. At each time step,
a pair is selected (asynchronous dynamics) and, if the pair is in the joint
state (+1+1) the price at each of the nodes is increased by one unit and the
state left unchanged. Similarly, if the pair is in the joint state (-1-1), the
price at each node is decreased by one unit and the state left unchanged. If the
joint state is (+1-1), it is assumed a deal takes place and the price associated
with each node is left unchanged, however the state of each node is reset randomly.
\par
The way the pair of nodes is selected needs further explanation. The first node, i
is selected at random. The route used to choose the second node is conditioned by
the way nodes are linked, ie the form of the network. If having chosen the first
spin i we insist that the second is either i+1 or i-1 i.e, a next neighbor, we
simulate a one-dimensional network. However if we also admit other nodes with
coordinates $i\pm g$ where $g>>1$ then we allow our node i to have 4 'near' neighbours (NN=4).
It can be shown \cite{ref4} that in such a case systems with short range interactions
behave as they possess NN/2=2 dimension. By admitting more well separated nodes,
such as $i\pm g'$ where $|g'|>>|g|$, we can increase the dimensionality further.
Thus for M additional nodes, $NN=2+M$ corresponding to a dimension $D=NN/2=1+M/2$.
If we allow our additional nodes to be selected with probability q where $0<q<1$
then we have a way of introducing fractional dimensions i.e., D = (1+q*M/2.) and
so change the value of the dimension continuously.
\section{Results}
\par
In this model prices differ on different spin locations not distinguished in
terms of geography. The differences arise from random changes in the local
node values. Local prices that are initially zero everywhere diverge as a
result of the model dynamics. The width of the price distributions may be
fitted by model parameters. Prices $p_i$ are always positive when evolving variables
correspond to $log(p_i)$ .
\par
Since the nodes are all equivalent, the price trajectory of the price on any
particular node may be thought of as one of a number of possible trajectories
that can occur on any node. Hence all the trajectories taken together comprise
an ensemble from which averages and expected values may be obtained.
\par
The graphs in figure 1 show the probability distribution of prices
as the dimension of the system changes from 1 to 2.2. For one dimension
the distribution is bimodal however as the dimension to 1.2, the bimodality
disappears and, by the time the dimension has reached 2.2 it becomes Gaussian.
\par
The model developed here has some similarity to the voter model \cite{ref5,ref6,ref7}
where the dynamics are due to random switching of spins at the edges of up and
down domains and there is a $50\%$ probability that a domain up will grow with time.
In our model, when the dimension is unity, the stationary state consists of two
large domains of opposite spin state. Each domain exhibits a price corresponding to
one of the bimodal regions in the price distribution. Continuing to increase the
number of links, and hence dimension of the system, enhances the possibility for
break up of large domains. Since the system is random, Gaussian distributions must
eventually appear.
\par
In Figure 2 we sketch details of the topology for both 1 and 2 dimensional networks.
In 1 dimension we see that the probability of domains being broken is comparable
to the probability of an increase or decrease. However, for 2 dimensions the
probability of breakup is much larger than the probability of either preservation
or an increase/decrease in size of the domains.
\par
We explored the evolution of the PDF for 1D further. In Fig.3 the evolution
of the PDF as the function of increasing number of iterations is shown.
One can see a monotonic increase of price in the peak and the height of
this peak when the network evolves in time. Fig. 4 shows more precisely
the increase of the price in the peak in time. In our calculations we
always are out of stationary state. We see this problem in Fig.5 in which
the plot of domain numbers in time decreases linearly with log-log axis.
The network evolves in time on a two domains state. In Fig. 6 we present
the dependence of price in the peak from network size. In the case of larger
systems the domains are larger so the price in the peak also.
\section{Risk}
\par
It is common in financial circles to equate risk with variance.
We make the same association and identify risk with the variance of
the different price distributions. The result is shown in figure 7.
\par
For small dimensions, the risk is relatively high. On a log log scale, it
falls quickly as the dimension increase and the shape of the distribution
becomes unimodal. From then on, up to a dimension of two the variance remains
more or less constant. As the dimension continues to increase, the risk falls.
Again there is a 'cusp' phenomenon as the dimension goes through integer values,
however the fall may be said to be roughly linear as the dimension increases to
these high values.
\par
From figure 1 we have seen the qualitative change that occurs as the dimension
of our system evolves from 1 to 2 and then to 3. For low dimensions few persons
are involved in making the deal and the associated risk is relatively high all
the more so because of the bimodality in the price distribution. As we increase
the number of buyers and sellers in the deal making process the price distribution
forms a Gaussian and the associated risk falls. The shape of the plot in Fig. 7 does
not change qualitatively with system size.
\par
The phenomenon discussed here can be observed with assets such as rare
objects or even cars in the third world where few people are involved in
the deal making process and where large price differences can be found.
\begin{figure}
\includegraphics[scale=0.35,angle=0]{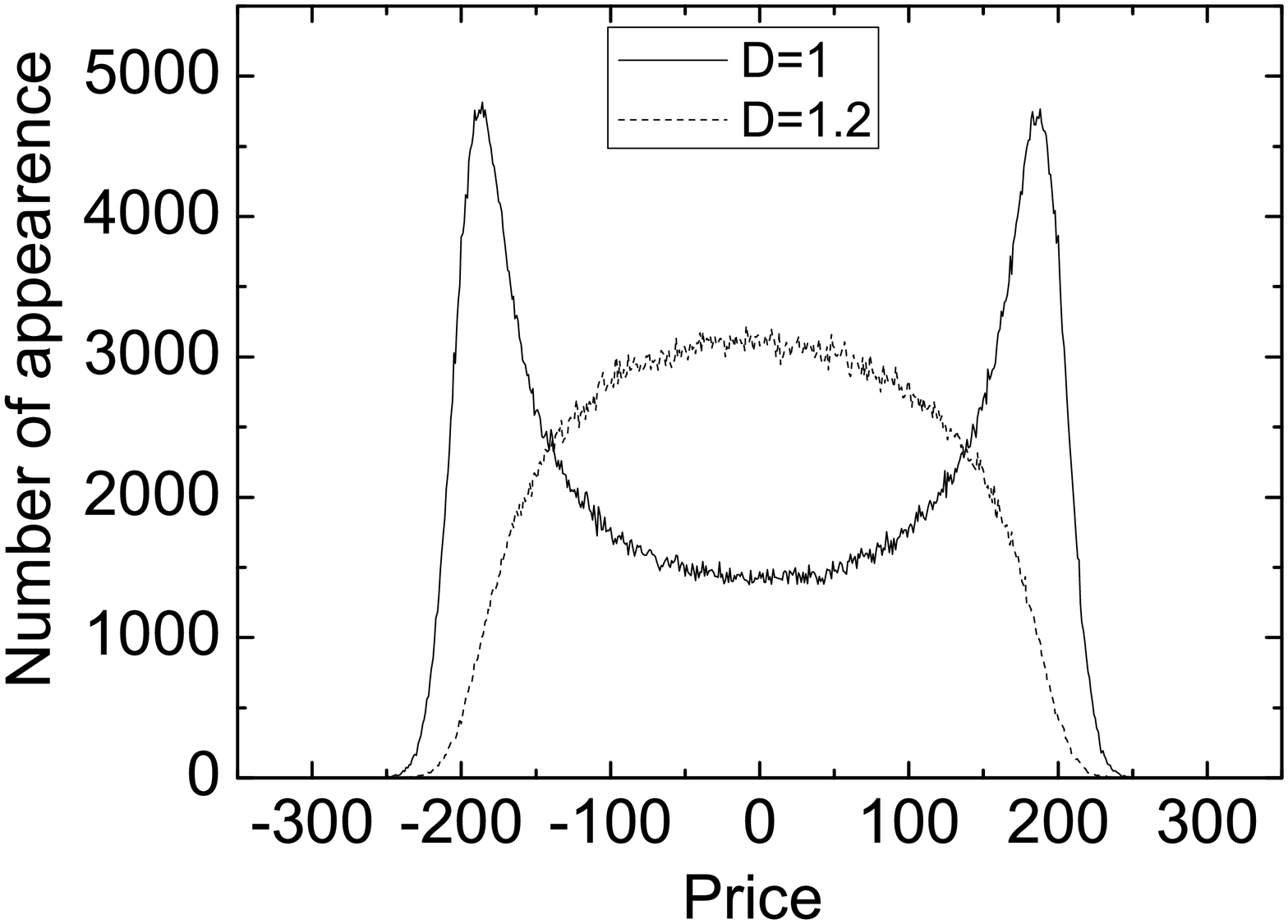}
\includegraphics[scale=0.35,angle=0]{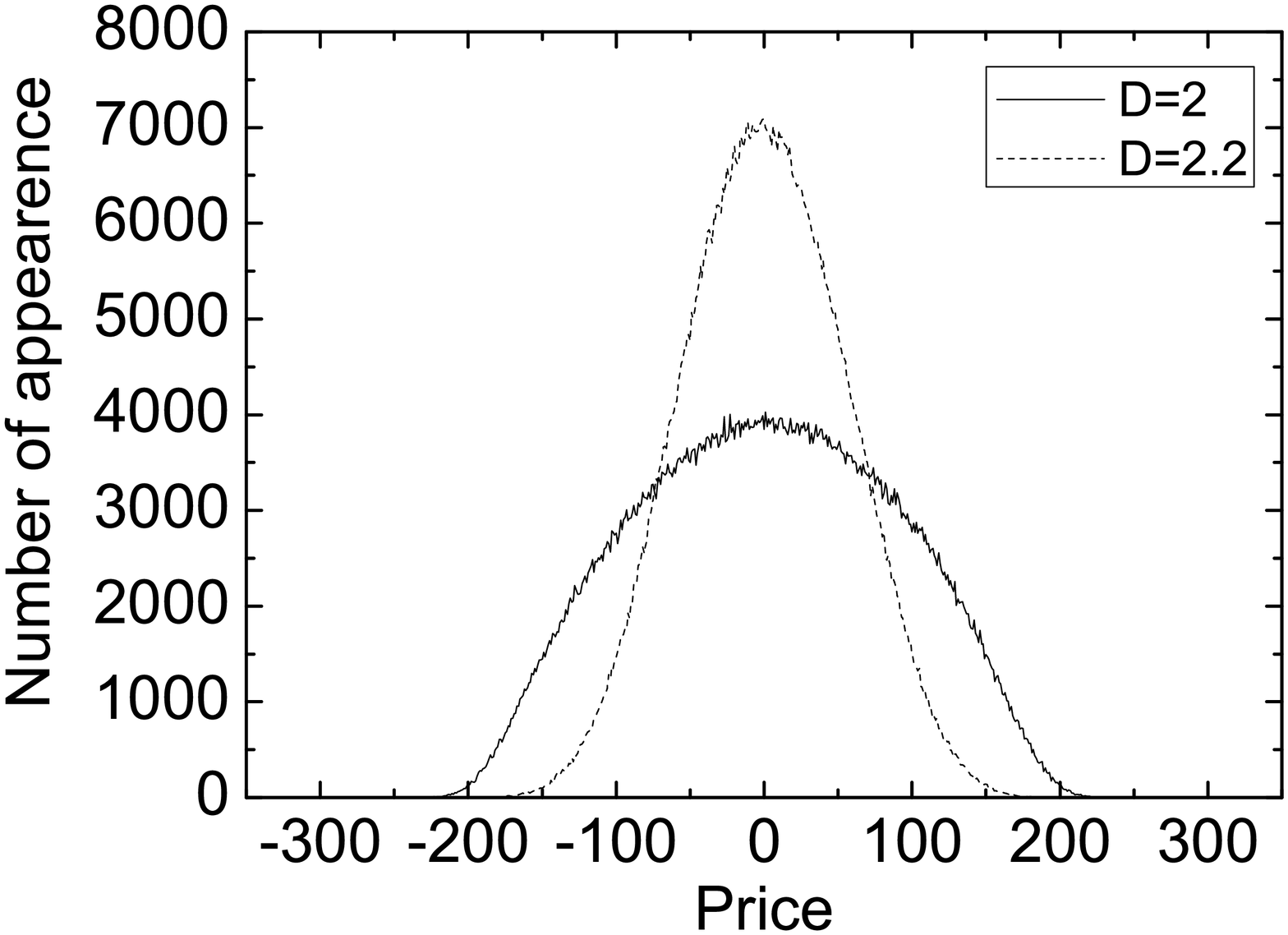}
\caption{The top curve shows the distribution at low dimensions and illustrates how,
as the dimension increases above 1, the bimodal character disappears. The lower curve
illustrates how Gaussian behaviour sets in as the dimension goes beyond 2.}
\end{figure}
\begin{figure}
\includegraphics[scale=0.5,angle=-90]{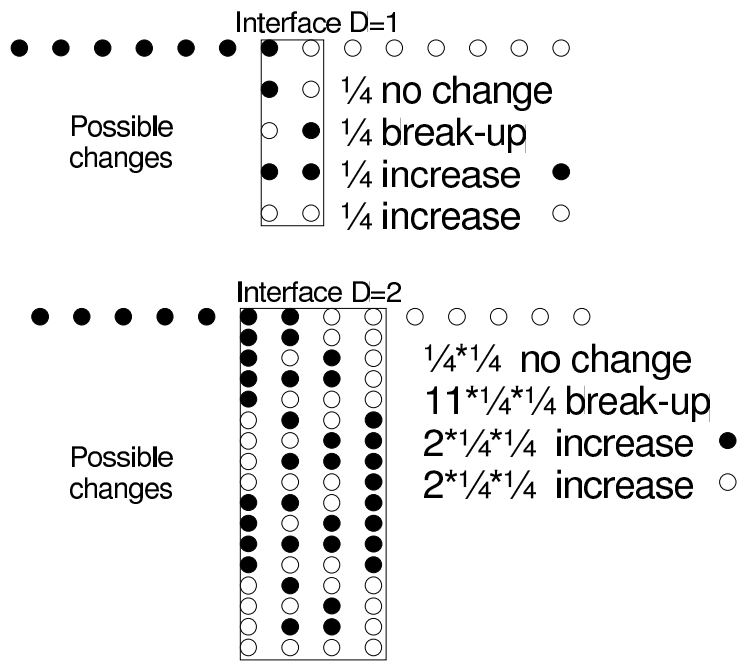}
\vspace{7cm}
\caption{Schematic diagram of the change in size of domains for 1 and 2
dimensional networks showing the probability associated with the various
different outcomes.}
\end{figure}
\begin{figure}
\includegraphics[scale=0.35,angle=0]{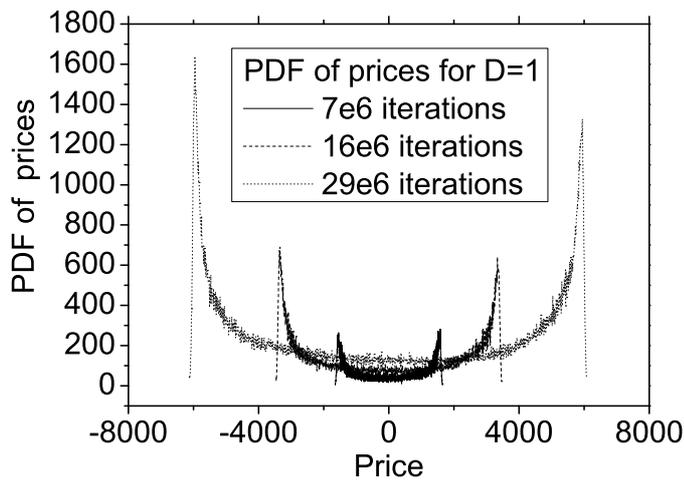}
\caption{The evolution of PDF in the case of increasing number of iteration.}
\end{figure}
\begin{figure}
\includegraphics[scale=0.35,angle=0]{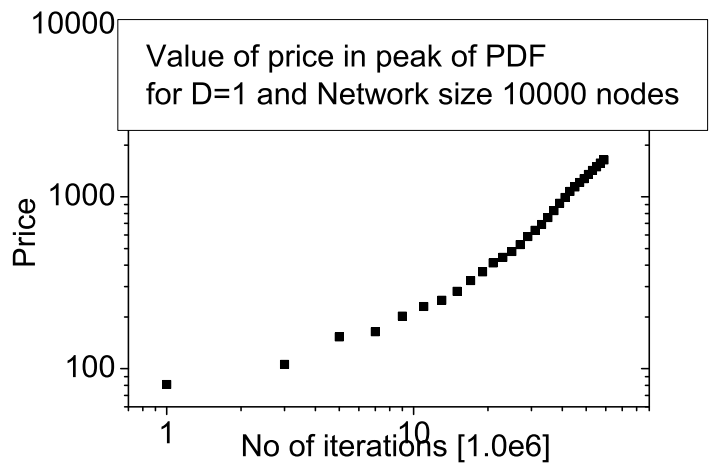}
\caption{The plot of evolution of peak price with increasing number of iteration. }
\end{figure}
\begin{figure}
\includegraphics[scale=0.35,angle=0]{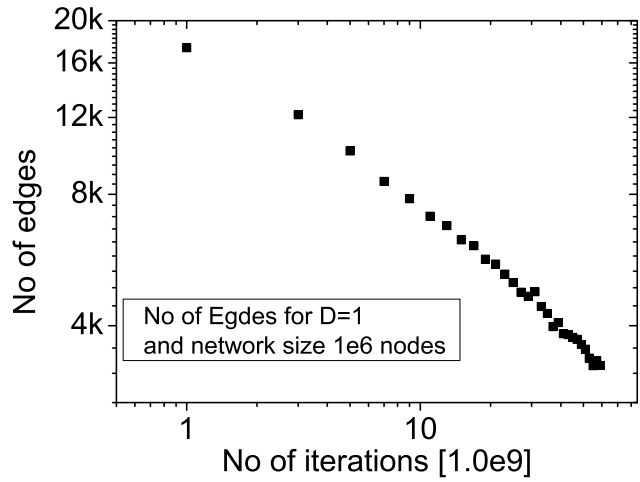}
\caption{The plot of evolution number of edges (number of domains) with
increasing number of iteration.}
\end{figure}
\begin{figure}
\includegraphics[scale=0.35,angle=0]{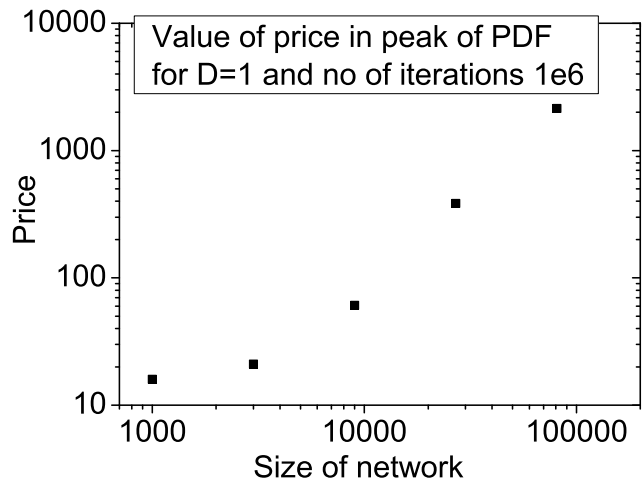}
\caption{The plot of peak price for different size of the network.}
\end{figure}
\begin{figure}
\includegraphics[scale=0.35,angle=0]{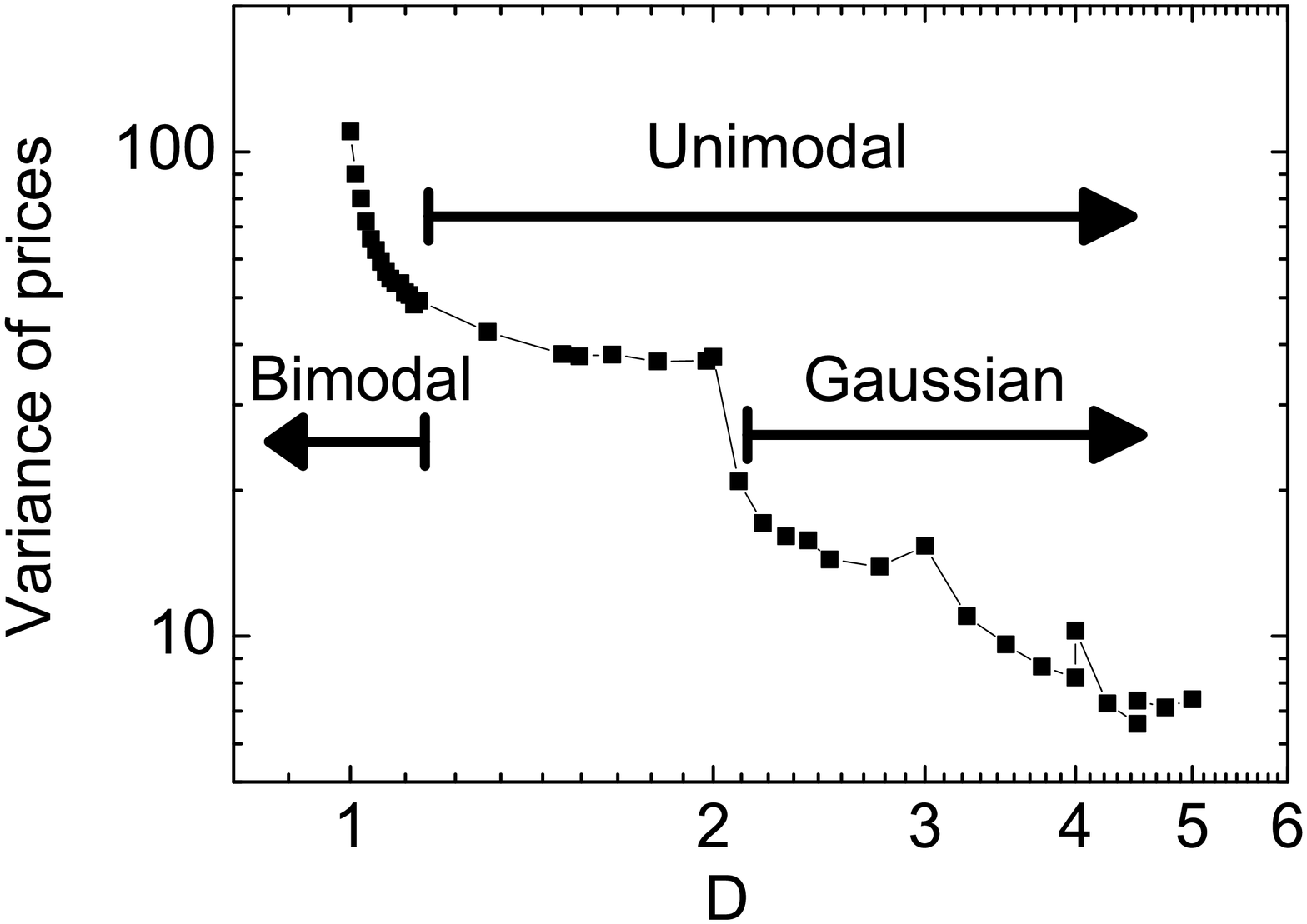}
\caption{Variance of prices (Risk) as a function of system dimension.}
\end{figure}
\section{Acknowledgment}
\par It is a  pleasure for us to dedicate this paper to Prof. Marcel
Ausloos and Prof. Dietrich Stauffer on  occasions of their 65th birthdays.
Authors acknowledge financial support from Polish Ministry of Science and Higher
Education,  Grant No. 134/E-365/SPB/COST/KN/DWM 105/2005-2007.

\end{document}